# Raman spectroscopy and planetary exploration: testing the ExoMars/RLS system at the Tabernas Desert (Spain)

Marco Veneranda[1], Guillermo Lopez-Reyes[1], Jose Antonio Manrique-Martinez[1], Aurelio Sanz-Arranz[1], Jesús Medina[1], Carlos Pérez[2], César Quintana[2], Andoni Moral[2], Jose A. Rodríguez[3], Jesús Zafra[3], Fernando Rull[1]

[1] *Department of Condensed Matter Physics, Crystallography and Mineralogy, Univ. of Valladolid, Spain. Ave. Francisco Vallés, 8, Boecillo, 47151 Spain. marco.veneranda.87@gmail.com*
[2] *National Institute for Aerospace Technology (INTA), Torrejón de Ardoz, Spain*
[3] *Systems Engineering for the Defense of Spain, S.A (ISDEFE), Madrid, Spain*

**Abstract**

*ExoFit* trials are field campaigns financed by ESA to test the *Rosalind Franklin* rover and to enhance collaboration practices between *ExoMars* working groups. During the first trial, a replicate of the *ExoMars* rover was remotely operated from Oxfordshire (United Kingdom) to perform a complex sequence of scientific operation at the Tabernas Desert (Spain). By following the ExoMars Reference Surface Mission (RSM), the rover investigated the Badlands subsoil and collected drill cores, whose analytical study was entrusted to the RLS (Raman Laser Spectrometer) team. The preliminary characterization of core samples was performed in-situ through the RLS Engineering and Qualification Model (EQM-2) and the Raman Demonstrator (RAD1), being this a new, portable emulator of the RLS. In-situ results where then complemented by laboratory analysis using the RLS ExoMars simulator and the commercial version of the Curiosity/CheMin XRD system. Raman data, obtained by closely simulating the operational constraints of the mission, successfully disclosed the mineralogical composition of the samples, reaching the detection of minor/trace phases that were not detected by XRD. More importantly, Raman analysis detected many organic functional groups, proving the presence of extremophile organisms in the arid sub-surface of the Tabernas Desert. In light of the forthcoming *ExoMars* mission, the results here presented proves that RLS could play a critical role in the characterization of Martian sub-surface environments and in the analytical detection of potential traces of live tracers.

**1 Introduction**

The ESA-Roscosmos/*ExoMars* mission is scheduled to be launched by summer 2022 [1,2]. The *Rosalind Franklin* rover will explore the Martian surface [3], investigate the subsoil stratigraphy [4], and drill cores down to a depth of 2 meters. Subsurface samples will be crushed by the Sample Preparation and Distribution System (SPDS) [5] and delivered to the analytical laboratory drawer (ALD) of the rover, where the Raman Laser Spectrometer (RLS) [6], MicrOmega [7] and MOMA [8] systems will collaborate in the search for signs of past and present life.

Being the first ESA rover ever operating in an extraterrestrial body, the *ExoMars* team needs to perfect the synergistic collaboration between instrument working groups and to practice with real mission issues. To do so, ESA organized the so-called *ExoMars*-like Field Testing (*ExoFit*) trials [9]. In brief the trials consisted in manoeuvring an emulator of the *Rosalind Franklin* rover (Charlie) in a Martian-like desertic landscape (Local Control Centre LCC) by following the instructions provided by the Remote Control Centre (RCC) team, which operates the entire mission simulation by only relying on the data returned from the rover [10].

The first *ExoFit* trial was carried out at the Tabernas Desert (Almeria, Spain), a terrestrial analogue site that offers a very similar landscape to what is expected to be found at the landing site of the *ExoMars* mission (*Oxia Planum*). During the two-weeks campaign which started on September 18, 2018, LCC and RCC teams carried out a complex sequence of scientific-technical operations by closely following the *ExoMars* Reference Surface Mission (RSM) [11]. By simulating the activities of 9 Martian sols, panoramic instruments onboard the *Rosalind Franklin* rover were used to investigate the site. After descending the landing platform, Charlie reached multiple areas of high scientific interest by avoiding rocks and further physical obstacles. At each area of interest, a combination of close-up images of the surface (CLUPI [12]) and Radar subsurface stratigraphy investigations (WISDOM [4]) were performed to identify the best spots for potential sample drilling. After drilling, subsurface materials were crushed and finally analyzed by the RLS team, who joined the *ExoFit* activities with a group of researchers and technical personnel from the University of Valladolid (UVa) and the National Institute for Aerospace Technology (INTA).

To simulate the potential scientific outcome of the RLS instrument, a set of Raman spectrometers developed in the framework of the *ExoMars* mission were used. Concretely, drill cores were analyzed at the LCC by using the Raman Demonstrator 1 (RAD1). RAD1 is a RLS portable prototype developed by the Erica research group (UVa) to perform in-situ analysis of terrestrial analogue sites [13]. By reproducing the diffraction grating, optical configuration and geometry of the RLS, this instrument offers qualitatively comparable data outcomes. Further in-situ analysis of Tabernas samples were carried out using the RLS engineering and qualification model 2 (EQM-2). As recently described elsewhere [14] the EQM-2 was assembled to demonstrate the ability of the RLS to 1) endure the stresses related to space travel and landing, 2) successfully operate under the harsh environmental conditions of Mars, and 3) fulfil the scientific capabilities required by the mission. RAD1 and EQM-2 results were then compared to laboratory spectra gathered through the RLS *ExoMars* simulator. This instrument was developed by the RLS team to obtain Raman data qualitatively equivalent to those of the RLS, while avoiding logistic problems related to the use of instruments qualified for space exploration [15,16]. As explained elsewhere, range of analysis, laser power output, spot of analysis and spectral resolution of RLS *ExoMars* Simulator closely resemble those of the RLS Flight Model (FM). Furthermore, this instrument integrates the same algorithms developed for the RLS to perform the automatic multi-point analysis of Martian samples, including SNR optimization, florescence quenching and acquisition parameters selection [17]. Besides Raman data, complementary analysis were performed by using an X-ray diffractometer (XRD), being this one of the most commonly used analytical tool for the mineralogical study of geological samples. XRD measurements were carried out through a commercial portable diffraction system based on the same technology of the CheMin instrument onboard the NASA/Curiosity rover [18,19]. Named Terra (Olympus). The quality of diffractometric data this system obtains from the study of terrestrial analogues is equivalent to those the CheMin is gathering from Martian rocks and soils [20].

By comparing Raman data with mineralogical information obtained from the commercial version of CheMin, one of the most successful analytical instruments ever deployed on Mars, this study seeks to 1) extrapolate reliable information about the potential scientific outcome of the RLS, 2) deepen the understanding about the role this instrument could play in the fulfilment of the mission objectives and, in a broader perspective, 3) quantify the impact that Raman spectroscopy could have in the mineralogical and exobiological exploration of extraterrestrial bodies.

**2 Materials and methods**

*2.1 Almeria Desert and its analogy with Oxia Planum*

Tabernas Badlands are located in a Neogene-Quaternary depression partially surrounded by the Betic cordillera system and filled with Neogene marine sediments consisting of calcareous and gypsiferous mudstone and calcareous sandstone [21,22]. Despite the arid conditions (235 mm of mean annual rainfall), the surface of Tabernas Desert is characterized by the proliferation of microorganisms [23]. Cyanobacteria and other autotrophic organisms increase the overall stability of the soil by binding mineral grains and giving rise to the formation of the so-called biological soil crust (BSC) [23,24]. It is well known that BSC increases nutrient availability in the uppermost layer of the soil, thus favouring the development of life forms of higher complexity (lichens) [25,26]. However, further studies proved that life proliferation drastically decreases as a function of depth (5 cm below the surface, the total organic carbon content drops by ≈90% [23]). Having this in mind, the study of Tabernas drill cores has a high scientific relevance for the *ExoMars* mission. Indeed, it offers the opportunity to assess the ability of Raman spectroscopy to reveal the mineralogical composition of sub-soil samples and, at the same time, to detect the potential presence of extremophiles biomarkers.

As shown in Figure 01a, the site chosen for the mission simulation is a flat area covered by poorly consolidated grains with mud patches and sporadic boulders. At the centre of the site is an elongated multi-layered ridge outcropping from the regolith ground. According to the visual inspection, the ridge is part of a turbidite sequence composed of alternating layers of sandstones and mudstones. As explained elsewhere, Badlands lithology is often characterized by high contents of phyllosilicate minerals generated from the alteration of primary rocks [27]. In the case of the Tabernas Desert, previous mineralogical studies of soils samples detected relevant contents of muscovite, paragonite and chlorite-smectite phases [28]. This mineralogical composition has a certain analogy with the landing site of the *ExoMars* mission. Indeed, according to orbital data presented elsewhere, *Oxia Planum* preserves vast phyllosilicate deposits [29,30]. As explained by Fornaro et al., Martian phyllosilicates are considered to be the optimal target for exobiological studies, since they are well acknowledged for being capable of hosting microbial life within their lamellar structure and preserving biomarkers [31].

The granulometry of Tabernas rocks and soils is another relevant aspect to take into consideration. As explained above, boulders and outcrops found at the simulated landing site are the results of the compaction of fine grains. Similarly, the particle size distribution of the regolith ground is dominated by grains of silt/clay-size. This evaluation fits with the results provided in previous studies, confirming that up to 80% of soils and sedimentary rocks at the Tabernas Desert are composed of silt-clay particles [21,23,32]. Similarly, *Oxia Planum* exhibits sedimentary rocks outcrops [33] as well as clay-rich formations of sedimentary origin [34]. Knowing that fine granulometries negatively affects the quality of Raman results (increase of the background level and peak width, decrease of SNR) [35], the Tabernas samples have served to evaluate to what extent the scientific capabilities of the RLS on Mars could be compromised by this parameter.

*2.2 Rover activity and sample collection*

The logistical and engineering challenges faced during the simulation, as well as the lessons learnt from the coordinated cooperation of instrument working groups is detailed in a specific work [36]. However, as summarized in Figure 01, it is important to highlight that, after a panoramic investigation of the area, the rover was commanded to approach the elongated turbidite ridge. Here, CLUPI and WISDOM systems were used to investigate the outcrop and to select the optimal sites for drilling. During the simulation, two cores were drilled by following the instructions provided by the RCC. The first core was collected from a poorly consolidated layer of the outcropping turbidite sequence. The drilling reached a depth of 25 cm, where a silt-sized powdered material was found. The second drill was carried out further along the ridge, on a sandstone material. In this case, the obtained sample comprised a mixtures of rock fragments and lose, fine-grained sediments. Considering that the analytical instrument onboard the *ExoMars* rover will investigate core samples in powder form, an agata mill was used to crush the rock fragments.

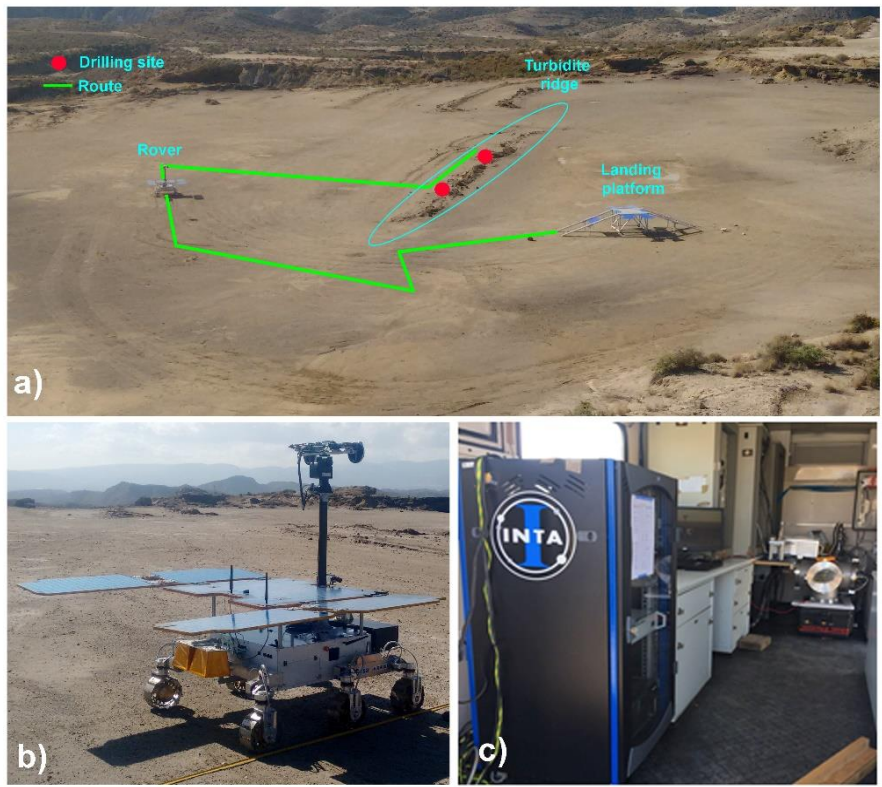

*Figure 01: a) selected site for the ExoFit trial b) close-up image of the Rosalind Franklin rover emulator used for the mission simulation; c) RLS team operating the RLS EQM-2 for the in-situ analysis of TDC1.*

*2.2 Instruments:*

*RLS EQM-2*

As is the case of the RLS flight model (FM), the engineering and qualification model (EQM-2) is made up by three main components: the spectrometer unit (SPU), the internal optical head (iOH) and the electronics control and excitation unit (ICEU). The ICEU is composed of three electronic boards (DC/DC power board, DPU Processor board and Detector FEE board) [6] and contains the excitation source, which is made of a Nd:YAG diode laser coupled to a second harmonic generator (SHG) crystal, thus emitting at 532 nm (power output between 10 and 20 mW, up to 7 mW on the sample). The ICEU also implements the ASW (Application SW) to provide the full control over the instrument to assess the possibility of implementing an autonomous operation: scientific operation, instrument management & monitoring, communications with the Rover and other functionalities as: focus capability, manage the timing and synchronization with Rover Vehicle Simulator (RVIS) and manage the scientific algorithms execution (parameter estimation algorithm, autofocus and fluorescence removal) Figure 01e. The iOH unit focuses the excitation laser into the samples and collects Raman scattering (Figure 01d). As described elsewhere, the iOH ensures a throughput >70%, a spectral collection range between 200 and 3800 cm-1, a focuser movement range of ± 1 mm and an excitation spot size of 50 ± 5 µm [37]. At the SPU, the Raman signal is collimated, diffracted and focused onto the CCD detector [38]. This unit was designed to fulfil the scientific requirements of the instrument (e.g. spectral resolution and SNR) and to ensure optimal performances in a thermal environment between -40ºC and 0ºC with low power consumption. To facilitate the in-situ analysis of drill cores, the EQM-2 was assembled in an INTA truck that served as mobile scientific laboratory. The vehicle was equipped with an electric generator to supply power to the EQM-2 and all the additional instrumentations necessary for its operation. As shown in Figure 01c, the SPU was incorporated in a vacuum chamber to reproduce the Martian atmospheric pressure.

RAD1

Complementary in situ analysis were performed with the RAD1 spectrometer. The instrument was assembled by the ERICA group (UVa) by using a commercial 532nm excitation laser, a high resolution TE Cooled CCD Array spectrometer (2168x512 pixels) and a high line density diffraction grating identical to the RLS instrument one (1800 lines per mm). The spectrometer has similar laser power output (7 mW on the sample), range of analysis (70-4200 $cm^{-1}$) and spectral resolution (6-10 $cm^{-1}$) to the RLS. Furthermore, the Raman probe was coupled to a microscope to achieve the 50 µm spot expected on the sample surface. The SPDS positioning system of the *Rosalind Franklin* rover was emulated with a micrometric linear positioner placed in the microscope tray. The microscope includes a camera (visible) to capture the sample before the analysis, somehow emulating the MicrOmega imaging, which was not possible during the simulation. For the spectral acquisition with RAD1, the spots of analysis were randomly chosen by the operator, while the measurement parameters were manually selected.

RLS ExoMars Simulator

Beside in-situ studies, further Raman analysis were carried out in the laboratory by using the RLS ExoMars simulator. As reported elsewhere this system allows to obtain spectra qualitatively comparable to those the RLS will gather on Mars, while avoiding the limitations imposed by the management of instrumentation developed for space exploration [15,39,40]. The instrument is composed of a BWN-532 excitation laser (B&WTek) emitting at 532 nm, a BTC162 high resolution TE Cooled CCD Array spectrometer (B&WTek) and an optical head with a long WD objective of 50x. To replicate RLS spectra, laser power on the sample (spot of 50 µm) war set to 7 mW and Raman analysis were collected over a wavenumber range that goes from 70 to 4200 $cm^{-1}$ with a mean spectral resolution of 8 $cm^{-1}$. Tabernas core were analyzed by replicating the RLS standard automated 39-point line analysis. Data collection was carried out by using a custom-made dedicated software coded with LabVIEW 2013 (National Instruments, USA).

Raman spectra were visualized an managed through the IDAT-SpectPro tool, which is the software programmed by the UVa-CSIC-CAB Associated Unit ERICA (on behalf of the RLS science team) for the reception, decodification, visualization and treatment of RLS data [41]. Spectra were finally interpreted by comparison with the RRUFF mineral spectra database [42].

XRD Terra

Raman results were compared with those obtained by using the Terra portable XRD instrument (Olympus). As anticipated in section 2, this system uses the same technology developed by NASA and onboard the MSL/Curiosity rover (CheMin instrument). The diffractometer consists of a CoKα excitation source (300kv, 300μA), a vibrating sample holder cell and a 2D peltier cooled CCD detector. X-ray diffraction is recorded in a range ranging from 5 to 55º 2θ with an average resolution of 0.25 -0.30º 2θ FWHM. Mineral identification was carried out using the BRUKER DIFFRAC.EVA software and the Powder Diffraction File-2 2002 (ICDD) database. Knowing that the Terra exitation source (CoKα) is different from those employed by the majority of laboratory diffractometers (CuKα), the diffractograms were converted to facilitate their comparison with the diffraction pattern contained in the PDF-2 database. To do so, the following equation was employed:

$$col(C)=114.59156*asin(L2/L1*sin(0.00872664*col(A)))$$

Where C is the value of the K-alpha emission of copper (1.54056 Ang) and A is the value of the k-alpha emission of cobalt (1,789 Ang)

**3 Results**

3.1 In situ analysis (Raman Spectroscopy)

The in-situ activities of the LCC team were drastically compromised during the first week of the trial due to heavy rains. The impact of bad weather on the usability of the simulated landing site and the rover itself made that the whole mission simulation had to be compressed within 9 sols (instead of 14). Accordingly, the timeframe established for the in-situ Raman characterization of drill cores was reduced to 1-hour per sample. Within the time constraints imposed by the mission simulation, RAD1 and EQM-2 instruments could collect between 6 and 8 spectra per sample, this being just a small fraction of the data-set the RLS will gather from Martian drill cores (20 to 39 analysis).

In spite of the mentioned limitation, the RAD1 study of TDC1 and TDC2 powdered cores enabled the detection of different mineral phases. As shown in Figure 02b, most of the spectra collected from both powdered materials returned a main peak at 1085 cm$^{-1}$ together with a weaker signal at 282 and 710 cm$^{-1}$. By comparing with the RRUFF database, these spectra perfectly fit with calcite patterns ($CaCO_3$). In addition to that, the characteristic signal of quartz ($SiO_2$) at 464 cm$^{-1}$ was clearly detected on both samples (Figure 02a). These results agree with previous mineralogical studies that confirmed calcite and quartz to be among the main mineral phases of Tabernas rocks and soils [43].

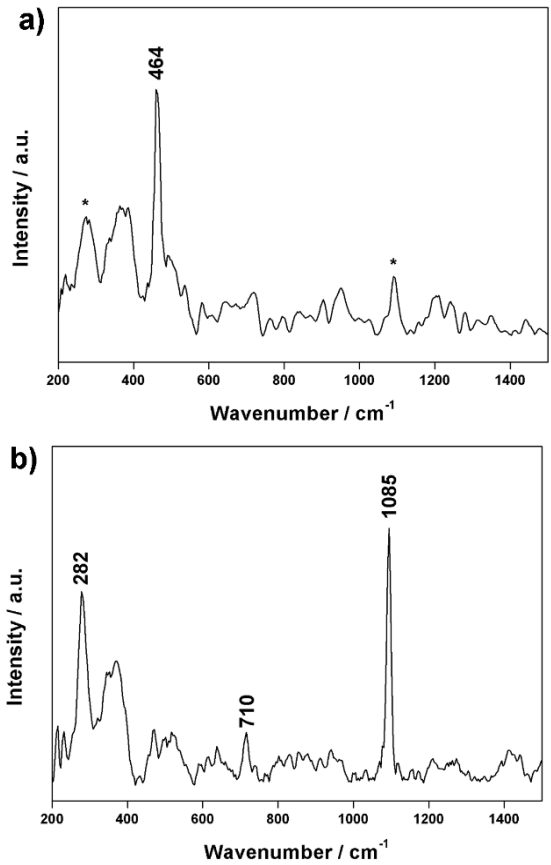

*Figure 02: RAD1 spectra of quartz and calcite. Raman signals proceeding from additional compounds are labelled with an asterisk.*

Compared to RAD1, the in-situ EQM-2 activities were limited to the first week of the trial due to the need of returning critical parts of the instrument to INTA for the integration of the RLS-FM. Within this time frame, only the TDC1 core could be analyzed. The powdered material was placed on the replicate of the *ExoMars* sample holder mounted below the iOH unit and, within the hour of analysis, a total of 7 spectra were collected. Considering that within the vacuum chamber, the SPU was not refrigerated and so the spectrometer detector was not working under Martian temperature conditions. For this reason, spectral resolution and SNR of RLS EQM-2 data was expected to be lower than RLS FM spectra from Mars. In spite of that, the RLS EQM-2 detected quartz (Figure 03) and calcite in the sample, thus confirming the results gathered from the RAD1 system.

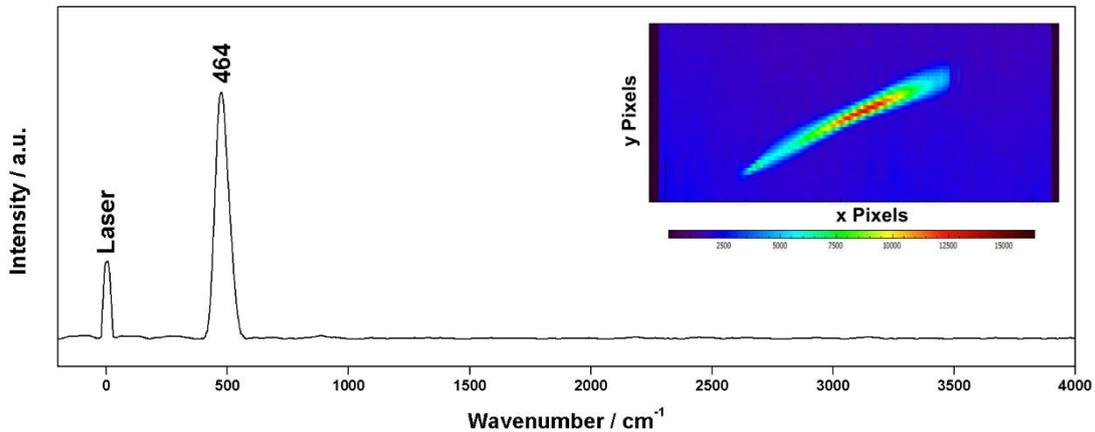

*Figure 03: RLS EQM-2 spectrum of quartz from sample TDC1. The raw CCD image collected by the SPU detector is also provided.*

As can be seen in Figure 04, some calcite spectra also displayed secondary peaks at 281 and 711 cm$^{-1}$. The detection of these additional signals on Martian samples will be of crucial importance for the *ExoMars* mission since they could serve to optimize the scientific outcome of Raman results. As is the case of many other mineral groups (e.g. nitrates and sulphates) the main peaks of carbonates is due to internal vibration modes (in this case, stretching vibration of the anionic group (CO$_3$)$^{2-}$ [44]. Knowing that this vibration is minimally affected by the cation (e.g. Ca$^{2+}$, Fe$^{2+}$, Mg$^{2+}$ and their mixtures), the variation in the main peak position of different carbonate minerals is very weak. On the contrary, the position of Raman peaks detected at lower wavelengths is strongly affected by the cations in the unit cell (external vibration modes). Therefore, the detection of secondary peaks is of crucial importance to discriminate among mineral phases on Mars, since they could be used to compensate the lower spectral resolution of Raman spectrometers developed and validated for space exploration (compared to state-of-the-art laboratory tools).

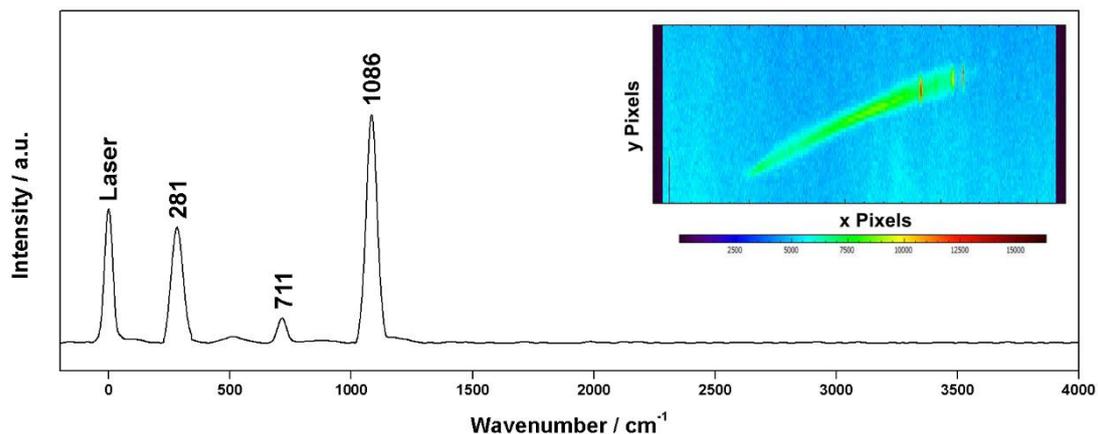

*Figure 04: RLS EQM-2 spectrum of calcite from sample TDC1. The raw CCD image collected by the SPU detector is also provided.*

3.2 Laboratory analysis:

3.2.1 Raman spectroscopy (RLS ExoMars Simulator)

The spectroscopic characterization of the two drilled cores was completed with laboratory analysis. Raman characterization of TDC1 and TDC2 samples was performed through RLS *ExoMars* Simulator by strictly following the operational constraint of the RLS. To do so, powdered samples were placed on a replicate of the *ExoMars* refillable sample holder and, after flattening, 39 spectra were automatically collected, being this the maximum number of analysis per sample established for a nominal measurement cycle. Laboratory results are summarized in Table 01. In brief, both drill cores displayed many spectra of quartz (Figure 05a) and calcite (Figure 05b). As displayed in Figure 05d, clear Raman peaks were often found at 264, 402, 695 and 3625 cm$^{-1}$, confirming the presence of muscovite in the two samples. Additionally, the multi-point analysis of core samples enabled the identification of additional mineral phases. For example, clear Raman peaks at 145, and 610 cm$^{-1}$ were detected in both cores, thus confirming the presence of anatase (TiO$_2$, Figure 05c). As displayed in Figure 05e, an additional titanium oxide (rutile, main peaks a 240, 445 and 610 cm$^{-1}$) was detected on sample ADC2. This mineral is commonly found on rocks submitted to high pressure and temperature, being an indicator of hydrothermal alteration processes or contact metamorphism. One of the spectra gathered from sample ADC2 displayed the characteristic doublet of feldspar minerals (478 and 508 cm$^{-1}$), together with several secondary peaks (165, 285, 407, 768, 805 and 1100 cm$^{-1}$). The vibrational profile of this spectra (Figure 05f) perfectly fits with albite patterns [45], this being the Na-rich end member of the plagioclase subgroup (NaAlSi$_3$O$_8$).

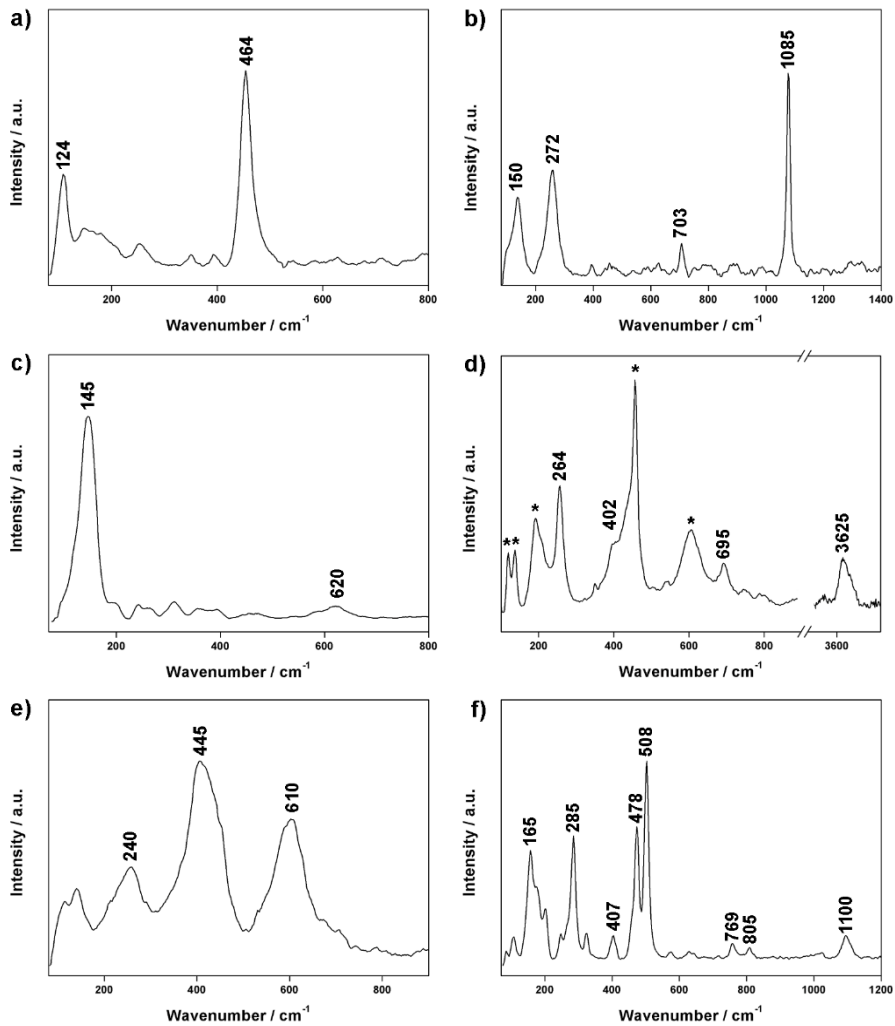

Figure 05: *Representative Raman spectra of inorganic compounds collected in the laboratory by using the RLS ExoMars Simulator: a) quartz, b) calcite, c) anatase, d) muscovite, e) rutile, f) plagioclase. Raman signals proceeding from additional compounds are labelled with an asterisk.*

Besides the detection of mineral phases, laboratory analysis of drilled cores enabled the identification of multiple biomarkers. As displayed in Figure 06, an intense peak at 1565 cm$^{-1}$ was fond on several TDC1 and TDC2 spectra. According to the detailed work published by Socrates, this signal can be associate to the excitation of -$NH_3$ and =$NH_2$ functional groups, as well as to the vibration of many different nitrogen-heterocyclic compounds, such as quinazoline ($C_8H_6N_2$), pyrroline ($C_4H_7N$) and imidazole ($C_3H_4N_2$) [46]. Regarding sample TDC2, an intense Raman signal was detected at 893 cm$^{-1}$. According to the information provided elsewhere, peaks around 890-900 cm$^{-1}$ are emitted by the C-C stretching vibrations of many different organic compounds [47]. In addition to those, many minor signals were found around 1300 cm$^{-1}$ and between 2350 and 2600 cm$^{-1}$. Although these signals could be attributed to the excitation of further organic functional groups, their low SNR could lead to misleading considerations.

The Raman characterization of biomarkers from Tabernas soils samples was successfully achieved in previous studies [48–50]. However, the high astrobiological relevance of the results here presented is due to the fact that 1) organic features were collected from subsoil samples (low content of organic carbon) rather than from biological soil crusts, 2) powdered materials were characterized through the RLS ExoMars Simulator by closely emulating the operational constraints of the RLS and 3) drilling sites were remotely selected by the RCC team, who was operating the rover by only relying on *ExoMars* instrumentation data.

The selection of a 532nm excitation source had a critical importance for the successful Raman detection of phyllosilicates and biomarkers. Indeed, it is well acknowledged that the intensity of the scattered light is proportional to the fourth power of the frequency, so that green lasers (532nm, as the one equipped by the RLS) is 4.7 and 16 times more efficient that red (785nm) and ultraviolet (1064nm) excitation sources, respectively. Furthermore, knowing that the optimal quantum efficiency (above 50%) of conventional CCD detectors is reached in the wavelength range between 400 and 680nm, the detection of Raman peaks at high wavelengths (above 1500 cm$^{-1}$ as is the case of hydration waters and many organic functional groups) is enhanced by the use of green lasers (Raman scattering region between 532 and 640nm) rather than red ones (scattering region between 785 and 1160nm). Even though the high energy of green lasers often leads to a stronger florescence background, the SNR of the spectrum can be increased by irradiating the spot of analysis for a period of time before data acquisition (quenching). As such, the RLS software integrates the algorithm necessary to automatically reduce the potential fluorescence emission of Martian samples. Those aspects make the RLS an optimal Raman spectrometer to investigate the water/geochemical environment of the Martian subsoil and to search for potential biomarkers.

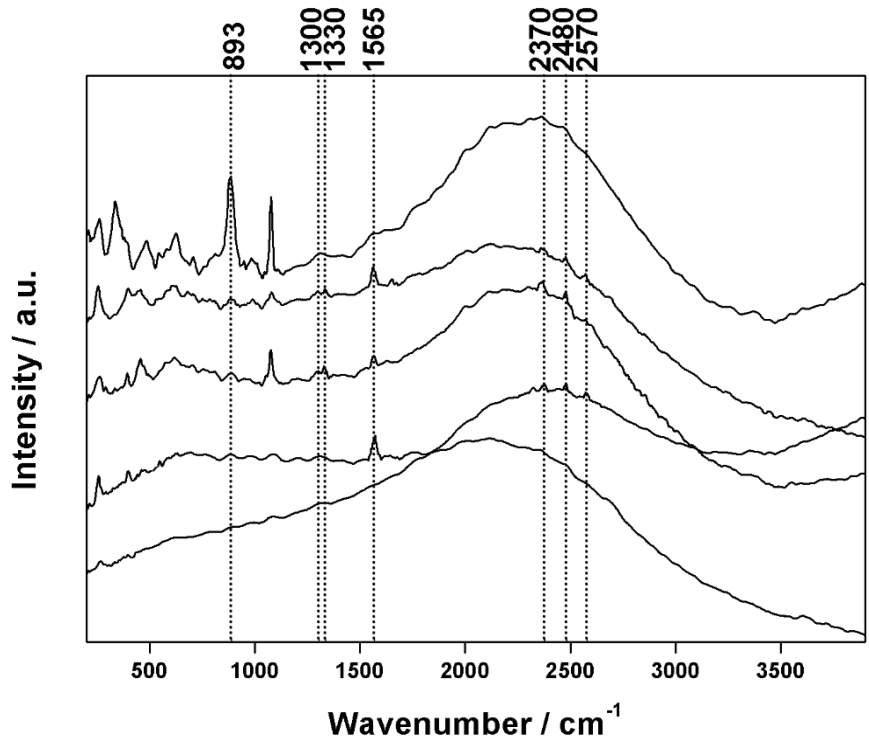

*Figure 06: Representative Raman spectra of organic compounds collected from TDC1 and TDC2 cores by using the RLS ExoMars Simulator*

3.2.2 X-ray diffractometry (Terra)

The diffractograms obtained from the two drill cores are shown in Figure 06. Compared to the results provided by conventional XRD systems, the detected diffractometric peaks display a higher FHMW value (full width at half maximum). This effect is not to be attributed to the characteristics of the samples, but rather to the novel design of the portable XRD system. Indeed, unlike the reflection geometry used by conventional XRD models with goniometers, Terra instrument features a transmission geometry where the X-ray beam pass through the powdered sample, which is positioned in a vibrating sample holder [20]. Despite the effect of peak broadening, the maximum intensity of the diffractometric signals does not vary in position, thus allowing the reliable identification of the detected mineral phases.

As can be seen in Figure 06, the two diffractograms can be almost perfectly overlapped, which indicates an equal mineralogical composition. Quartz (based on 20.90 and 26.65 2θ values) and calcite (29.45 and 39.45 2θ) were found to be the main mineral phases of both cores, thus confirming in-situ (RAD1 and RLS EQM-2) results. The two diffractograms also revealed additional diffractions that could be assigned to phyllosilicate minerals. In detail, the peaks detected at 8.85 and 19.95 2θ are characteristic of muscovite, while the signals at 12.50 and 25.05 confirmed the presence of chlorite. Besides mineral phases, the two diffractograms did not show any background, which suggests the absence of amorphous inorganic material.

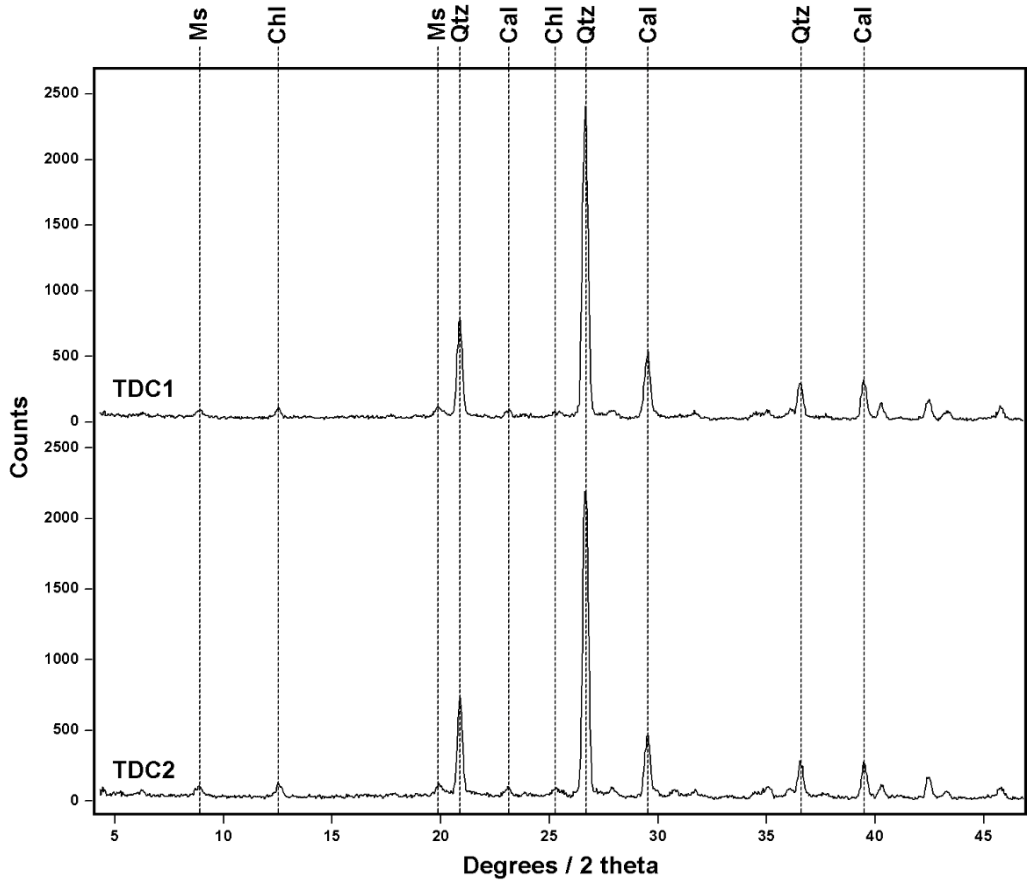

*Figure 07: Diffractograms obtained from the analysis of samples TDC1 and TDC2, revealing the presence of chlorite (Chl), muscovite (Ms), quartz (Qtz) and calcite (Cal).*

From the summary provided in Table 01, it can be noted that the RLS ExoMars Simulator missed the detection of chlorite, which was found by XRD as minor phase of both drilled cores. This could be due to the fact that the weak Raman scattering exhibited by most phyllosilicates can be covered by the signal of additional minerals. This confirms that phyllosilicates are hardly detected by Raman spectrometers when representing the minor component of heterogeneous mineral mixtures. On the other hand, laboratory Raman results obtained by simulating the operational constraints of the RLS allowed to identify additional compounds that were below the detection limit of XRD, as is the case of anatase (in both samples), rutile and plagioclase (TDC1). This is a very important result since, as inferred from previous works [51–53], help demonstrating that the analytical strategy based on the multipoint Raman analysis of powdered samples could effectively help detecting minor or trace compounds, from which important inferences about the geological and environmental evolution of Mars can be inferred.

*Table 01: Overview of the mineralogical results gathered from the use of spectroscopic and diffractometric systems. * RLS-EQM-2 analysis of sample TDC2 are not available.*

| SAMPLE | Quartz | Calcite | Feldspar | Anatase | Rutile | Muscovite | Chlorite |
|---|---|---|---|---|---|---|---|
| TDC1 | 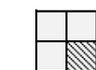 | 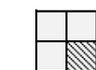 | | 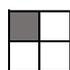 | | 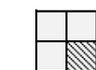 | 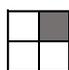 |
| TDC2* | | | | | | | |

Legend: Analysis not performed / RLS ExoMars / XRD / RAD1 / RLS-EQM2

## 5 Conclusion

Through the *ExoFit* trial, the RLS team was able to extrapolate important inferences about the potential scientific outcome of the Raman spectrometer onboard the *Rosalind Franklin* rover. For the first time, the RLS EQM-2 was applied for the in-situ characterization of core samples drilled during an *ExoMars* mission simulation. Even though only 7 spectra could be collected (over the 20-39 of a nominal mission operation), the RLS EQM-2 enabled the detection of the main mineral phases of the sample (quartz and calcite). EQM-2 data perfectly agreed with those collected through the RAD1 system, thus demonstrating that this portable RLS simulator can be a valuable tool for the in-situ investigation of *ExoMars*-related terrestrial analogues sites. Laboratory Raman analysis, performed through the RLS *ExoMars* Simulator by closely emulating the operation constraints of the mission, helped demonstrating that the multipoint Raman analysis of powdered samples is the optimal analytical strategy to maximize the detection of minor and trace compounds. In addition to identifying additional mineral phases (from which, information about the geological and environmental evolution of Mars could be extrapolated), this study proved that RLS could effectively detect biomarkers. Knowing that the Tabernas Desert has many mineralogical and environmental similarities with *Oxia Planum*, the detection of many organic functional groups on both drilled cores is an extremely important result for the *ExoMars* team as it confirms the ability of the RLS to detect potential extremophilic microorganisms colonizing the subsurface of Martian-like environments. In a broader perspective, this study also strengthens the idea that Raman spectroscopy can be the ideal technique to apply to all space-exploration missions that, beyond characterizing the mineralogy of extra-terrestrial bodies, aim to identify the possible presence of life tracers.

**Acknowledgements:**

This work is financed through the European Research Council in the H2020-COMPET-2015 programme (grant 687302) and the Ministry of Economy and Competitiveness (MINECO, grants ESP2017-87690-C3-1-R and PID2019-107442RB-C31).